\documentclass[a4paper,11pt]{article}
\usepackage{jcappub, bm, color} 
\usepackage{amssymb,amsfonts,slashed,amsthm,amsmath,graphicx, soul}
\usepackage[caption=false]{subfig}
\title{Tunneling wave function of the universe}

\author{Alexander Vilenkin}
\author{and Masaki Yamada}

\affiliation{Institute of Cosmology, Department of Physics and Astronomy, 
Tufts University, Medford, MA  02155, USA}

\emailAdd{vilenkin@cosmos.phy.tufts.edu}
\emailAdd{Masaki.Yamada@tufts.edu}

\newcommand{\beq}{\begin{eqnarray}} 
\newcommand{\eeq}{\end{eqnarray}}

\def\({\left(}
\def\){\right)}
\def\[{\left[}
\def\]{\right]}

\def\nn{\nonumber \\}

\def\lmk{\left(}
\def\rmk{\right)}
\def\lkk{\left[}
\def\rkk{\right]}

\newcommand{\eq}[1]{Eq.~(\ref{#1})}
\newcommand{\bel}[1] {\begin{equation}\label{#1}}
\newcommand{\beal}[1] {\begin{eqnarray}\label{#1}}
\newcommand{\be}{\begin{equation}}
\newcommand{\ee}{\end{equation}}
\newcommand{\bea}{\begin{array}} 
\newcommand{\eea}{\end{array}}
\newcommand{\abs}[1]{\left\vert#1\right\vert}

\def\del{\partial}

\abstract{
The tunneling wave function of the universe is investigated in a minisuperspace framework of a de Sitter universe with a quantum scalar field, treated as a perturbation.  We consider three different approaches to defining the tunneling wave function: (1) tunneling boundary conditions in superspace, (2) Lorentzian path integral, and (3) quantum tunneling from initial universe of a vanishing size.  We show that the superspace approach requires Robin boundary conditions for the scalar field modes, the path integral approach requires adding an appropriate boundary term to the scalar field action, and the initial universe approach requires the initial quantum state of the scalar field to be Euclidean vacuum.  We find that all three approaches yield identical wave functions and that scalar field fluctuations are well behaved, contrary to earlier claims in the literature.
}

\begin{document}

\maketitle
\flushbottom

\section{Introduction}

Inflationary spacetimes are known to be past-incomplete \cite{Borde:2001nh}.  This indicates that such spacetimes must have a past boundary and raises the question of what determined the initial conditions at that boundary.  Even though it may not be essential for observational predictions of inflationary models, this is an important question of principle, and without resolving it the inflationary cosmology remains incomplete.  Perhaps the most promising approach to this problem is based on quantum cosmology, which suggests that a spatially compact universe can spontaneously nucleate out of `nothing', where `nothing' refers to a state with no classical space, time and matter \cite{Vilenkin:1982de,Hartle:1983ai,Linde:1983mx,Rubakov:1984bh,Vilenkin:1984wp,Zeldovich:1984vk}.

In quantum cosmology the universe is described by a wave function $\Psi(g,\phi)$, which is defined on the space of all possible 3-geometries $(g)$ and all matter field configurations $(\phi)$, called superspace.  The role of the Schr\"{o}dinger equation for $\Psi$ is played by the Wheeler-DeWitt equation \cite{DeWitt:1967yk}
\beq
{\cal H}\Psi = 0,
\label{WDW1}
\eeq
where ${\cal H}$ is the Hamiltonian operator.  This is a functional differential equation in superspace.

In ordinary quantum mechanics, the wave function of a system is found by solving the Schr\"{o}dinger equation with boundary conditions determined by the physical setup external to the system.  But since there is nothing external to the universe, it appears that the boundary conditions for the Wheeler-DeWitt equation should be postulated as an independent physical law.  Several possible forms of this law have been discussed in the literature; the most developed proposals are the Hartle-Hawking and the tunneling boundary conditions.\footnote{Alternative boundary conditions have been introduced by DeWitt's \cite{DeWitt:1967yk} and Linde \cite{Linde:1983mx}.  These proposals, however, have been discussed only in the context of one-dimensional minisuperspace models, and no attempt has been made so far to extend them to full superspace.}
  
The tunneling boundary condition has been discussed in detail in Refs.~\cite{Vilenkin:1986cy,Vilenkin:1987kf,Vilenkin:1994rn}.  Roughly, it requires that $\psi$ should include only outgoing waves at the boundary of superspace, except for the part of the boundary corresponding to vanishing 3-geometries.  This is supplemented by the regularity condition, requiring that $\psi$ remains finite everywhere, including the boundaries of superspace,
\beq
\abs{\Psi(g,\phi)}<\infty.
\label{regularity}
\eeq  
Thus, the probability flux enters superspace through 3-geometries of vanishing size and leaves it through the rest of the boundary, corresponding to singular or infinitely large universes.  The resulting wave function can be interpreted as describing a universe originating at zero size, that is, from `nothing'.  It was conjectured in \cite{Vilenkin:1984wp} that the same wave function can be expressed as a path integral over Lorentzian histories interpolating between a vanishing 3-geometry and a given configuration in superspace, 
\be
\Psi_T=\int_\emptyset^{(g,\phi)} {\cal D} g {\cal D} \phi \, e^{iS},
\label{psiT}
\ee
where $S$ is the action.  However, the equivalence of the two definitions has not yet been generally demonstrated.
  
The Hartle-Hawking (HH) wave function \cite{Hartle:1983ai} is defined as a path integral over compact Euclidean `histories' bounded by given 3-geometry and matter field configuration,
\be
\Psi_{HH}=\int^{(g,\phi)} {\cal D} g {\cal D} \phi \, e^{-S_E},
\label{psiHH}
\ee
where $S_E$ is the Euclidean action obtained by the standard Wick rotation $t\to -i\tau$.  This wave function has also been interpreted as describing quantum nucleation from nothing.  The gravitational part of the Euclidean action $S_E$ is unbounded from below, and as it stands the integral (\ref{psiHH}) is divergent.  One can attempt to fix the problem by additional contour rotations, extending the path integral to complex metrics \cite{Halliwell:1988ik, Halliwell:1989dy}.  However, the space of complex metrics is very large, and no obvious choice of integration contour suggests itself as the preferred one.

This is where things stood until the last year, when Feldbrugge, Lehners and Turok (FLT) reinvigorated the field with a new approach to Lorentzian quantum cosmology \cite{Feldbrugge:2017kzv}.  Working in minisuperspace framework, they showed that the path integrals (\ref{psiT}), (\ref{psiHH}) can be rigorously defined with the aid of the Picard-Lefschetz theory.  They first applied this method to de Sitter minisuperspace model and found that in this case the Euclidean path integral cannot be made convergent by any deformation of the lapse integration contour, while the Lorentzian path integral is well defined and gives the tunneling wave function, as it was claimed in \cite{Vilenkin:1984wp,Vilenkin:1994rn}.  (See Refs.~\cite{Halliwell:1988ik, Halliwell:1989dy, Brown:1990iv} for related earlier work.)  In the following papers \cite{Feldbrugge:2017fcc,Feldbrugge:2018gin} FLT considered perturbative minisuperspace, with the gravitational wave field $\phi$ added as a small perturbation.  Here, their results for the tunneling wave function were far less reassuring.  They found that this wave function predicts a runaway instability, where the probability of quantum fluctuations of the field $\phi$ grows with their amplitude.  Similar claims about instability of the tunneling proposal have also been made in the earlier literature \cite{Halliwell:1989dy}.

FLT work has led Diaz Dorronsoro {\it et al.} \cite{DiazDorronsoro:2017hti,DiazDorronsoro:2018wro} to further develop the HH proposal.  They studied de Sitter plus scalar field and Bianchi-IX models and showed that for specific choices of the lapse integration contour in the complex plane the wave function is well defined and exhibits the qualitative behavior expected of $\Psi_{HH}$.%
\footnote{
Dispute about the behavior of these wave functions beyond perturbation theory is still ongoing~\cite{Feldbrugge:2017mbc,Feldbrugge:2018gin,DiazDorronsoro:2018wro}, but here we focus exclusively on perturbative superspace.
}
In our view, however, the basic criticism against this approach still remains: the HH proposal is incomplete without a choice of a complex integration contour in the path integral (\ref{psiHH}).  Some general requirements to this contour have been given in Ref.~\cite{Halliwell:1989dy}, but it is not clear that they can always be satisfied or what contour should be used in models admitting a number of choices that satisfy the requirements.%

In the present paper we focus mostly on the path integral formulation of the tunneling proposal.  We show that the field fluctuations in the wave function (\ref{psiT}) are well behaved if the action $S$ is supplemented with a suitable boundary term.  In the next section we review the calculation of $\Psi_T$ using the tunneling boundary conditions in perturbative superspace and give an alternative formulation of the boundary conditions, which is more suitable for our purposes here.   
In Section 3 we introduce into the action a boundary term, which is appropriate for these boundary conditions, and show that the resulting path integral coincides with the wave function obtained using the tunneling boundary conditions.  We also propose a physical interpretation of the boundary term in terms of the initial scalar field wave function in a tunneling universe.  Our conclusions are summarized in Section 4.

In the main text of the paper we consider a massive conformally coupled scalar field.  The case of a minimally coupled field, which in the massless case is equivalent to that of gravitational waves, is discussed in the Appendix.  Throughout this paper, we use the reduced Planck units: $8 \pi G = 1$.

\section{Tunneling boundary conditions}
\label{sec:wavefunction}

\subsection{The model}

We consider a de Sitter minisuperspace model with a conformally coupled massive scalar field, 
where the metric is assumed to be homogeneous, isotropic and closed: 
\beq
 d s^2 = a(\eta)^2 ( -N^2 d \eta^2 + d \Omega^2_3). 
 \label{metric}
\eeq
Here, $N$ is the lapse function, $a$ is the scale factor, $\eta$ is the conformal time, and $d\Omega_3^2$ is the metric on a unit 3-sphere. 
In this section the lapse function is irrelevant and is set to be unity.

The action for this model is given by 
\beq
S=\int \sqrt{-{g^{(4)}}} \, d^4 x \lmk \frac{R}{2}- 3 H^2 \rmk + S_m
+S_B ,
\label{action}
\\
S_m = \int \sqrt{-g^{(4)}} \, d^4 x \left[ - \frac{1}{2}(\nabla\phi)^2-\frac{1}{2}m^2\phi^2 -\frac{\xi}{2}R\phi^2\right], 
\eeq
where $g^{(4)}$ is the determinant of the metric, $H={\rm const}$ is the de Sitter expansion rate, $\xi = 1/6$ is the conformal coupling, $S_m$ is the matter part of the action, and $S_B$ is the boundary term.  
The boundary term is not relevant in this section and will be specified in Sec.~\ref{sec:PI}. 
The case of a minimally coupled field (i.e., $\xi = 0$) is discussed in the Appendix. 

We expand the field $\phi$ as
\beq
\phi(x,t)=\frac{1}{a(t)} \sum \chi_n(t)Q_n(x),
\eeq
\beq
\int Q_n Q_{n'} d\Omega_3 = \delta_{n n'},
\eeq
where $Q_{nlm}(x)$ are suitably normalized spherical harmonics and we have suppressed the indices $l,m$ for brevity.  
The superspace of this model is an infinite-dimensional space $\{a,\chi_n\}$.
The wave function of the Universe obeys the Wheeler-DeWitt (WDW) equation: 
\beq
 \lkk \frac{\hbar^2}{24 \pi^2} \frac{\del^2}{\del a^2} - 6\pi^2 V(a) - \sum_n {\cal H}_n \rkk \Psi (a, \chi_n) = 0, 
 \label{WDW}
\eeq
where 
\beq
 V(a) = a^2 - H^2 a^4 
 \label{Va}
 \\
 {\cal H}_n \equiv \frac{\hbar^2}{2} \frac{\del^2}{\del \chi_n^2} - \frac12 \omega_n^2 (a) \chi_n^2 
 \label{H_n}
 \\
 \omega_n^2(a)=n^2 + m^2 a^2 
 \label{omega_n}
\eeq
with $n=1,2,...$.  In this paper we disregard the ambiguity of factor ordering. 

With the modes $\chi_n$ treated as small perturbations, a solution of Eq.~(\ref{WDW}) can be expressed as a superposition of terms of the form \cite{Halliwell:1984eu,Wada:1986uy, Vachaspati:1988as}
\beq
 \Psi (a, \chi_n) = A \exp \lkk - \frac{12\pi^2}{\hbar} S(a) - \frac{1}{2\hbar} \sum_n R_n (a) \chi_n^2 \rkk, 
 \label{wavefunction}
\eeq
where $A$ is a normalization constant.  Substituting this in (\ref{WDW}), we neglect terms ${\cal O}(\chi_n^4)$.  We also
use the WKB approximation for $S(a)$ and neglect terms ${\cal O}(\hbar)$. 
Then the WDW equation is rewritten as 
\beq
 \lmk \frac{dS}{da}\rmk^2 - V(a) = 0,
 \label{EoM:Sa}
 \\
 \lmk \frac{dS}{da} \rmk \lmk \frac{dR_n}{da} \rmk - R_n^2 + \omega_n^2 (a) = 0.
 \label{EoMSn}
\eeq

\subsection{Tunneling boundary conditions}

The wave function has different behavior in the classically allowed $(V(a)<0)$ and classically forbidden $(V(a)>0)$ regions. 
For the tunneling wave function, we require that $\Psi$ includes only an outgoing wave in $a$ at large $a$. 
Thus, for $V (a) < 0$ the wave function should be given by Eq.~(\ref{wavefunction}) with 
\beq
 S(a) = i  \int_{a_*}^{a} \sqrt{-V(a')} d a' + C ,
 \label{Sa}
\eeq
where $a_*=H^{-1}$ is the classical turning point defined by $V(a_*) = 0$ and $C={\rm const}$. 

In the under-barrier region $V(a)>0$, the wave function can be expressed as
\beq
\Psi(a,\chi_n)
&=&A_+ \exp\left[-12\pi^2 S^+(a)-\frac{1}{2} \sum_n R_n^+(a)\chi_n^2\right] 
\nn
&&+A_- \exp\left[- 12\pi^2 S^-(a)-\frac{1}{2} \sum_n R_n^-(a)\chi_n^2\right],
\label{psi}
\eeq
where
\beq
 S^\pm(a) = \mp \int_{a}^{a_*} \sqrt{V(a')} d a' + C .
 \label{Spm}
\eeq
The $A_+$ and $A_-$ terms correspond respectively to decreasing and growing wave functions.
We have chosen the integration constant $C$ in Eqs.~(\ref{Sa}),(\ref{Spm}) to be the same, so that $S(a_*)=S^\pm(a_*)$.   With this choice the coefficients $A$ and $A_\pm$, which can be determined by the WKB connection formulas, have comparable magnitude, $A_+\sim A_- \sim A$.  Their precise form, which was found in Ref.~\cite{Vilenkin:1984wp}, will not be important for our discussion here.

For $a<a_*$ it will be convenient to introduce a Euclidean conformal time variable $\tau$ via 
\beq
 \frac{da}{d \tau} \equiv 
 \left\{ 
 \bea{ll}
  \sqrt{V(a)} ~~~~{\rm for}~~ \tau < \tau_*
 \vspace{0.3cm}\\
 - \sqrt{V(a)}  ~~~~{\rm for}~~ \tau_* < \tau 
 \eea
 \right., 
 \label{dadeta}
\eeq
where the threshold $\tau_*$ is defined by $a (\tau_*) = a_*$. 
This can be solved as 
\beq
a(\tau)= (H\cosh \tau)^{-1}, 
\label{atau}
\eeq
where the turning point $a_*$ corresponds to $\tau_*=0$, and $a=0$ corresponds to ${\tau} \to \pm\infty$. 
This scale factor describes a Euclidean 4-sphere, which is an analytic continuation of de Sitter spacetime.
$a(\tau)$ in Eq.~(\ref{atau}) is an even function of $\tau$, so each value of $a<a_*$ corresponds to two values $\tau_\pm$ with $\tau_+(a)=-\tau_-(a)$.  We shall set $\tau_+(a)<0$.  

It follows from Eqs.~(\ref{Spm}) and (\ref{dadeta}) that 
\beq
 \frac{d S^\pm}{d \tau} = V(\tau) . 
\eeq
Then we can set
\beq
S^\pm = \int_{-\infty}^{\tau_\pm} V(\tau)d\tau.
\label{Spmtau}
\eeq
This corresponds to setting the integration constant $C$ in (\ref{Spm}) to
\beq
C=\int_{-\infty}^{\tau_*} V(\tau)d\tau =  \int_{0}^{a_*} \sqrt{V(a')} d a' .
\label{C}
\eeq
The actions $S_-$ $(S_+)$ then correspond to histories that do (do not) traverse the mid-section $a_*$ of the Euclidean 4-sphere as they go from $a=0$ to a given value $a$.

We now turn to Eq.~(\ref{EoMSn}) for the functions $R_n^\pm(a)$: 
\beq
 \lmk \frac{dS^\pm}{da} \rmk \lmk \frac{dR_n^\pm}{da} \rmk =\left(R_n^\pm\right)^2-\omega_n^2,
\label{Riccati}
\eeq
or 
\beq
 \frac{dR_n^\pm}{d \tau}=\left(R_n^\pm\right)^2-\omega_n^2 (\tau). 
\label{Riccati2}
\eeq
The matching conditions at $a=a_*$ require that $R_n^+(\tau_*)=R_n^-(\tau_*)$ \cite{Vachaspati:1988as,Hong:2003pe}. 
Since the functions $R_n^\pm$ satisfy a first-order differential equation of the same form and have the same value at $\tau=\tau_*$, they can be represented by a single function $R_n(\tau)$ with $\tau$ taken to be $\tau^+(a)$ $(\tau^-(a))$ for $R^+_n$ $(R^-_n)$.

Eq.~(\ref{Riccati2}) is a Riccati equation.  It can be reduced to a linear equation by the substitution 
\beq
 R_n(\tau) = - \frac{1}{\nu_n} \frac{d \nu_n}{d \tau},
\label{R_npm}
\eeq
where the mode functions $\nu_n(\tau)$ satisfy 
\beq
 \frac{d^2 \nu_n}{d \tau^2} - \omega_n^2 \nu_n =0.
\label{nueq}
\eeq

To impose the regularity condition (\ref{regularity}), we require that
\beq
{\rm Re} R_n^\pm (a)>0.
\label{regularity2}
\eeq
This ensures that $|\Psi|$ decreases as the amplitudes $\chi_n$ are increased, so that scalar field fluctuations are suppressed.
Strictly speaking, (\ref{regularity}) does not follow from (\ref{regularity2}), since our approximation breaks down at large values of $\chi_n$, but this seems the best one can do in a perturbative superspace model.  The condition (\ref{regularity2}) is non-local, in the sense that the functions $R_n^\pm(a)$ depend on the form of the potential $V(a)$ everywhere under the barrier.  We will now show that it can be replaced by equivalent local boundary conditions for $\nu_n$ at $\tau\to -\infty$.

It has been shown in \cite{Hong:2003pe} that the condition (\ref{regularity2}) is satisfied for $R_n^+(a)$ at $a<a_*$ and for $R_n(a)$ in the classically allowed range $a>a_*$, provided that it is satisfied at the turning point $a=a_*$. However, this is not automatic for $R_n^-(a)$ under the barrier, and the regularity condition gets violated at small $a$, unless we adopt a special choice of boundary conditions at $a\to 0$.  Specifically, it was shown in \cite{Hong:2003pe} that the regularity condition is satisfied by $R_n^-$ everywhere under the barrier if it is satisfied at $a\to 0$, or $\tau\to \infty$.  (Here we choose the branch $\tau^-(a)>0$ appropriate for the functions $R_n^-(\tau)$.)

To explore the behavior of the mode functions $\nu_n(\tau)$ at $a \to 0$ (or $\tau\to \pm\infty$), we can replace $\omega_n^2 \approx n^2$ in Eq.~(\ref{nueq}).  Then the solution is
\beq
\nu_n (\tau)\approx A_n e^{- n \tau}+B_n e^{n \tau}, 
\label{nusolution}
\eeq
and
\beq
 R_n(\tau) \approx n \frac{A_n-B_n e^{ 2 n \tau}}{A_n+B_n e^{ 2 n \tau}}.
\eeq
It is now easy to see that $R_n (\tau \to \infty) = -n <0$, unless we set $B_n=0$, or $\nu_n (\tau \to \infty)\propto \exp(-n \tau)$.
This corresponds to the boundary condition 
\beq
 \frac{d\nu_n}{d \tau} = - n \nu_n ~~~ (\tau \to  \infty).
\label{bc-}
\eeq  
Note that for a massless field the solutions (\ref{nusolution}) are exact in the entire range $-\infty<\tau<\infty$. In this case $R_n^\pm(a)=R_n(a)=n$ and the $\chi_n$- and $a$-dependent parts of the wave function factorize -- as they should for a conformally invariant field.

To find the boundary conditions for $\nu_n$ at $\tau\to -\infty$, we note that since Eq.~(\ref{nueq}) is symmetric with respect to the replacement $\tau\to -\tau$, the mode function $\nu_n(\tau)$ can be expressed in terms of a symmetric function $g_{sn}(\tau)$ and an antisymmetric function $g_{an}(\tau)$ as
\beq
\nu_n(\tau)=A_n \left[g_{sn}(\tau)-g_{an}(\tau)\right] +B_n\left[g_{sn}(\tau)+g_{an}(\tau)\right],
\label{nu_n sym}
\eeq
where $g_{sn}(-\tau)=g_{sn}(\tau)$, $g_{an}(-\tau)=-g_{an}(\tau)$, and $g_{sn}(\tau\to\infty)\approx \cosh(n\tau)$, $g_{an}(\tau\to\infty)\approx \sinh (n\tau)$.  It then follows from (\ref{nu_n sym}) that for $\tau\to -\infty$
\beq
 \nu_n (\tau) &\approx& A_n \lmk g_{sn} (-\tau) + g_{an} (-\tau) \rmk 
 \\
 &\approx& A_n e^{-n\tau},
\eeq
and the boundary condition is the same as (\ref{bc-}),
\beq
 \frac{d\nu_n}{d \tau} = -n \nu_n ~~~ (\tau \to - \infty).
\label{bc+}
\eeq  
After imposing the matching conditions at $a = a_*$ to determine the mode function in the classically allowed range, 
one finds that this choice of the mode functions corresponds to the Bunch-Davies vacuum state~\cite{Hong:2003pe}.

If the boundary condition (\ref{bc+}) is enforced at $\tau\to -\infty$, then, according to the results of \cite{Hong:2003pe}, $R_n^-(a)$ satisfy the regularity condition in the range $0<a\leq a_*$.  On the other hand, the matching conditions at $a=a_*$ require that \cite{Vachaspati:1988as} $R_n(a_*)=R_n^+(a_*)=R_n^-(a_*)$, and it follows from the analysis in \cite{Hong:2003pe} that the regularity condition is satisfied by $R_n^+(a)$ and $R_n(a)$ as well.  Thus we conclude that the boundary conditions (\ref{bc+}) are sufficient to ensure that the regularity condition is satisfied in the entire range $0<a<\infty$.  Combined with the outgoing wave condition, these boundary conditions uniquely determine the tunneling wave function in our model.  We show in the Appendix that the same conclusions apply to the de Sitter model with a minimally coupled scalar field.

The mode functions $\nu_n (\tau)$ selected by the boundary condition (\ref{bc+}) diverge at $\tau \to - \infty$.  This may look worrisome, but we note that $R_n^\pm$ and the wave function (\ref{psi}) are well behaved at $a\to 0$.  We therefore see no reason to require finiteness of $\nu_n(\tau\to -\infty)$ in the tunneling approach.  We shall return to this issue in Sec.~3.

\section{Path-integral formulation}
\label{sec:PI}

\subsection{de Sitter minisuperspace}

In the path integral formalism, the transition amplitude from an initial state $(g_0, \phi_0)$ to a final state $(g_1, \phi_1)$
can be symbolically written as
\beq
G(g_0,\phi_0;g_1,\phi_1)=\int_{(g_0,\phi_0)}^{(g_1,\phi_1)} {\cal D} g {\cal D} \phi \, e^{iS}.
\label{G}
\eeq
We will be interested in the limit where the initial geometry shrinks to a point.

We first consider the leading-order homogeneous de Sitter minisuperspace model.  
In the gauge where the lapse function is $N={\rm const}$, the path integral in (\ref{G}) reduces to 
\beq
G^{(0)}(0; a_1)=\int_0^\infty dN \int {\cal D}a ~ e^{iS^{(0)}[a,N]},
\label{PsiG}
\eeq
where 
\beq
 S^{(0)}[a,N] = 6 \pi^2 \int_{-\infty}^{\eta_1} \lkk - \frac{\dot{a}^2}{N} + N V(a) \rkk d \eta
\label{SdS}
\eeq
is the action for the de Sitter model with metric (\ref{metric}).  
Note that our starting point is a purely Lorentzian path integral and the lapse integration is performed over a semi-infinite range $N>0$.  The latter condition ensures that we include only histories where $\eta_1$ occurs later than $\eta=-\infty$ and not the geometrically identical histories related to them by $\eta\to -\eta$.  This is the choice adopted in Refs.~\cite{Teitelboim:1983fk, Feldbrugge:2017kzv, Vilenkin:1994rn}.  With this choice, $G(a_0,a_1)$ is a Green's function of the Wheeler-DeWitt operator satisfying
\beq 
{\cal H} G(a_0;a_1)=-i\delta (a_1-a_0).
\eeq
However, in the limit of $a_0\to 0$, $G(0;a_1)$ is a solution of the WDW equation in the entire range $0<a_1<\infty$.  

The path integral over $a$ in (\ref{PsiG}) was first evaluated by Halliwell and Louko \cite{Halliwell:1988ik}.  They noticed that the analysis can be greatly simplified by introducing a new time coordinate $t$, which is related to $\eta$ by $d\eta=a^{-2}(t)dt$ and which can be chosen to vary between 0 and 1.  Then the metric takes the form
\beq
ds^2=-\frac{N^2}{q(t)}dt^2+q (t) d\Omega_3^2,
\eeq
where $q=a^2$, and the action (\ref{SdS}) becomes
\beq
S^{(0)}[q,N]=6\pi^2\int_0^1 \left[-\frac{{\dot q}^2}{4N} +N(1-H^2 q)\right] dt .
\label{S0q}
\eeq
This action is quadratic in $q$, so the path integral can be evaluated exactly.

The classical equation of motion for $q(t)$ obtained by minimizing the action (\ref{S0q}) is
\beq
{\ddot q}=2N^2 H^2,
\eeq
its solution satisfying the boundary conditions $q(0)=0$ and $q(1)=a_1^2$ is
\beq
 q (t) = H^2 N^2 t^2 + \lmk a_1^2 - H^2 N^2 \rmk t, 
 \label{qt}
\eeq
and the action (\ref{S0q}) for this solution is given by
\beq
S^{(0)}(a_1,N)=6\pi^2\left(N^3\frac{H^4}{12}+N\left(1-\frac{1}{2}H^2 a_1^2\right) -\frac{a_1^4}{4N}\right).
\label{S0q2}
\eeq
The propagator (\ref{PsiG}) can now be expressed as \cite{Halliwell:1988ik}  
\beq
G^{(0)}(0; a_1)=\int_{0^+}^\infty \frac{dN}{N^{1/2}} e^{iS^{(0)}(a_1,N)},
\label{Nintegral}
\eeq
where we have omitted an overall numerical factor. 

The remaining integration over $N$ can be performed using the saddle point approximation.  The relevant saddle points and the steepest descent contours in the complex $N$-plane can be found using Picard-Lefschetz theory. 
The action (\ref{S0q2}) generally has four saddle points, which can be found from $\partial S^{(0)}/\partial N =0$.  In the under-barrier region $a_1<a_*$, the corresponding values of $N$ are purely imaginary and the relevant saddle points are located at \cite{Halliwell:1988ik,Feldbrugge:2017kzv}
\beq
 N^\pm = \frac{i}{H^2} \lmk 1 \mp \sqrt{1-H^2 a_1^2} \rmk. 
 \label{Npm}
\eeq
The corresponding actions are
\beq
S^{(0)}(a_1,N^\pm)=12\pi^2 i S^\pm(a_1),
\label{S0pm}
\eeq
where $S^+(a)$ and $S^-(a)$ are given by Eqs.~(\ref{Spm}),(\ref{C}) and correspond respectively to Euclidean paths that do and do not traverse the mid-section of the 4-sphere.  The pre-factors of the $\exp(iS^{(0)}(a_1,N^\pm))$ terms have comparable magnitudes at $a\sim a_*$.  In fact, it can be verified~\cite{Brown:1990iv} that they differ by a factor $i/2$, as they should for the tunneling wave function.  

At this point it will be convenient to switch back to the time variable $\eta$, which is related to $t$ as
\beq
t=\frac{2i}{H^2 N} \frac{1}{e^{2iN\eta}+1}.
\eeq
It can be verified that the classical solution (\ref{qt}) is then given by $a(\eta)=1/H\cosh(-iN\eta)$, which is the same as (\ref{atau}), up to the normalization of $\eta$.  (Note that the Euclidean time $\tau$ that we used in the Section 2 is related to $\eta$ via $i \tau = N \eta$.)  The values $t=0$ (where $a=0$) and $t=1$ (where $a=a_1$) correspond respectively to $\eta\to -\infty$ and  
\beq
 \eta_1^\pm = 
 \frac{i}{2N^\pm} \ln \lmk \frac{-H^2 (N^\pm)^2}{a_1^2} \rmk . 
\eeq
One can check that $\eta_1^+ < 0$ and $\eta_1^- > 0$.  The actions (\ref{S0pm}) can now be expressed as
\beq
 S^{(0), N^\pm} = 12 \pi^2 \int_{-\infty}^{\eta_1^\pm} V(a) N^\pm d \eta, 
\eeq
which is equivalent to (\ref{Spmtau}).

In the classically allowed range $V(a_1) < 0$ one finds that only one saddle point contributes \cite{Halliwell:1988ik,Feldbrugge:2017kzv},
\beq
 N = \frac{1}{H^2} \lmk i + \sqrt{H^2 a_1^2 -1} \rmk , 
 \eeq
and the transition amplitude (\ref{Nintegral}) is
\beq
G^{(0)}(0,a_1)\propto \exp\left(-12\pi^2 S(a_1)\right)
\eeq
with $S(a)$ given by Eqs.~(\ref{Sa}),(\ref{C}).  This describes a wave traveling in the positive $a_1$-direction, as expected for the tunneling wave function.  Thus, we can identify
\beq
G^{(0)}(0;a)=\Psi_T(a).
\eeq

\subsection{The problem with perturbations}
\label{sec:PI1}

We now consider the wave function for scalar field fluctuations $\chi_n$.  Assuming that the backreaction of fluctuations on the metric is negligible, the scale factor and the lapse parameter can be treated as background variables.  For definiteness we shall consider the under-barrier wave function with $a_1<a_*$.  Then the path integral in Eq.(\ref{psiT}) reduces to
\beq
\Psi_T(a_1,\phi_{n1})=\sum_{r=\pm} A_r e^{iS^{(0)}(a_1,N^r)} \prod_n \int {\cal D}\phi_n e^{iS_n [\phi_n; N^r]}
\label{Psiaphi}
\eeq
where
\beq
 S_n[\phi_n; N] = \frac12 \int_{\eta_0}^{\eta_1} d \eta \left(\frac{a^2}{N}{\dot\phi}_n^2 -Na^2(n^2-1)\phi_n^2 -Na^4 \left(m^2+\frac{1}{6}R\right)\phi_n^2\right) + S_{Bn}
 \label{Sn2}
\eeq
with $\eta_0\to -\infty$.
The integration is taken over histories where $\phi_n(\eta) \equiv \chi_n(\eta)/a(\eta)$ have specified values $\phi_{n1}$ at $\eta=\eta_1$, with suitable boundary conditions for $\phi_n$ at $\eta\to -\infty$.  The boundary term in (\ref{Sn2}) is usually not included; we shall discuss it in the next subsection.  From now on we omit the superscript $r=\pm$ for notational simplicity.

The path integral in (\ref{Psiaphi}) is again determined by the history $\phi_n(\eta)$ satisfying the classical equation of motion: 
\beq
\frac{1}{N^2}\left({\ddot\phi}_n+2\frac{\dot a}{a} {\dot\phi}_n\right) + (n^2-1)\phi_n + (m^2+2H^2)a^2\phi_n=0
 \label{EoM2}
\eeq
Disregarding the boundary term for the moment, the action is then given by 
\beq
 S_n = \frac{a_1^2}{2 N} \phi_n (\eta_1) \dot{\phi}_n (\eta_1) - \frac{a_0^2}{2 N} \phi_n (\eta_0) \dot{\phi}_n (\eta_0). 
 \label{solution2}
\eeq

In the limit of $\eta \to - \infty$, $a(\eta)\approx 2H^{-1}e^{-i N\eta}$ and the solution to the equation of motion is approximated to be 
\beq
 \phi_n (\eta) \approx A_n e^{i (n+1) N \eta} + B_ne^{-i (n-1) N \eta} . 
 \label{phi_n}
\eeq
Since ${\rm Im} N > 0$ for both saddle points in Eq.~(\ref{Npm}), 
the second term of $S_n$ diverges in the limit of $\eta_0 \to - \infty$ unless we take $A_n = 0$. 
This seems to suggest that we should set $A_n=0$ in order to make the action finite.  
With this choice, the $\phi_n$-dependent part of the wave function becomes (in the regime where the approximation (\ref{phi_n}) is applicable)
\beq
\psi_n(\phi_{n1})\propto e^{iS_n}=\exp\left(ia_1^2 \frac{\phi_{n1}{\dot\phi}_{n1}}{2N}\right)=\exp\left(\frac{n-1}{2} a_1^2 \phi_{n1}^2\right),
\eeq
which has the obvious problem that the wave function grows with increasing amplitude of the fluctuations.  One can check that the problem persists in the classically allowed range $a>a_*$.
This is the basis for numerous claims made in the literature that the tunneling wave function predicts an unstable runaway behavior of the fluctuation modes \cite{Halliwell:1989dy,Feldbrugge:2017fcc,DiazDorronsoro:2018wro}.  We shall see, however, that the problem can be resolved by inclusion of a suitable boundary term.

\subsection{Boundary terms}
\label{sec:PI2-1}

The boundary term $S_B$ in the action (\ref{action}) should be selected in such a way that boundary contributions obtained after varying the action and integrating by parts vanish, once the boundary conditions are imposed.  The form of the boundary term, of course, depends on the choice of boundary conditions.  The choice adopted in most of the literature on quantum cosmology is Dirichlet boundary conditions, with the 3-metric and the scalar field specified at the boundary.
The corresponding boundary action is
\beq
S_B=S_{GH} + \xi \int_{\cal B} \sqrt{-g^{(3)}} \, d^3 y K \phi^2 ,
\label{SB1}
\eeq
where $S_{GH}$ is the Gibbons-Hawking term, 
$\xi$ is the scalar field coupling to the curvature, 
$g^{(3)}$ is the determinant of the induced metric on the boundary ${\cal B}$, 
$y^a$ are the coordinates on the boundary, 
and $K$ is the extrinsic curvature of the boundary.  The second term in (\ref{SB1}) is absent for a minimally coupled field, but in our case $\xi=1/6$ and it has to be included in order for the variation of the action with respect to the metric not to give any uncompensated boundary terms \cite{Barvinsky:1995dp}.

The Dirichlet boundary conditions are appropriate for the Hartle-Hawking wave function,\footnote{This is because the mode functions $\chi_n(\eta)$ in the Hartle-Hawking approach are required to have specified values at $\eta=\eta_1$ and to satisfy $\chi_n(\eta\to -\infty)=0$.  This is sometimes justified by the requirement that the scalar field action should be finite.  We note, however, that the logic here is somewhat circular: the finiteness of the action depends on the choice of the boundary term, which in turn depends on boundary conditions.} but for the tunneling wave function one needs to take a different approach.  The spacetimes included in the path integral for $\Psi_T$ have two boundaries: the upper boundary ${\cal B}_1$ $(\eta=\eta_1)$ with specified values of $a_1$ and $\chi_{n1}$ and a lower boundary ${\cal B}_0$ where $a\to 0$ $(\eta_0\to -\infty)$.  
According to Eq.~(\ref{bc+}), the histories $\chi_n(\eta)=a(\eta)\phi_n(\eta)$ for the tunneling wave function should satisfy the Robin boundary condition
\beq
\frac{1}{N}\frac{d\chi_n}{d\eta}=in\chi_n ~~~ (\eta\to -\infty),
 \label{bcchi}
\eeq
where we have used the relation $i\tau=N\eta$.  We will now show that a suitable choice of the boundary terms in this case is
\beq
S_B=S_{GH} +\frac{1}{2\pi^2} \sum_n \int_{{\cal B}_0} \sqrt{-g^{(3)}} \, d^3 y \left( \xi K -\frac{1}{2}h_n \right) \phi_n^2
+\frac{1}{2\pi^2} \sum_n \int_{{\cal B}_1} \sqrt{-g^{(3)}} \, d^3 y \xi K \phi_n^2 ,
\nn\label{SB}
\eeq
where $h_n$ are parameters to be determined.

Variation of the action with respect to $\phi_n$ gives
\beq
&& 2\pi^2 \delta S= -\int \sqrt{-g^{(4)}} \, d^4 x\, \delta\phi_n \left(-\nabla^2+ \xi R+m^2\right)\phi_n 
\nn
&& +\int_{{\cal B}_0} \sqrt{-g^{(3)}} \, d^3 y \delta\phi_n \left(\partial_\perp \phi_n +2\xi K\phi_n -h_n\phi_n\right)
+\int_{{\cal B}_1} \sqrt{-g^{(3)}} \, d^3 y \delta\phi_n \left(\partial_\perp \phi_n + 2\xi K\phi_n \right).\nn
\label{variation}
\eeq
Here, $\partial_\perp$ is the derivative in the direction of outer normal to the boundary.  For our metric (\ref{metric}) it is given by
\beq
 \partial_\perp = \pm \frac{1}{N a} \frac{d}{ d \eta} ,
\label{dperp}
\eeq
where the upper and lower signs correspond to upper and lower boundaries, respectively.
The boundary term in (\ref{variation}) vanishes on the upper boundary ${\cal B}_1$, where $\phi_n$ are fixed, while on the lower boundary ${\cal B}_0$ we will impose the boundary conditions
\beq
\partial_\perp \phi_n + 2\xi K\phi_n -h_n\phi_n=0.
\label{bc}
\eeq
Noticing that
\beq
K = \frac{\partial_\perp V_B}{V_B}=3\frac{\partial_\perp a}{a},
\eeq
where $V_B = 2\pi^2 a^3$ is the boundary volume, we can express (\ref{bc}) as
\beq
\partial_\perp \chi_n-h_n\chi_n =0 ,
\label{chibc}
\eeq
where we have used $\xi=1/6$.
This coincides with (\ref{bcchi}) if we set
\beq
h_n = -i n a^{-1}. 
\label{BC}
\eeq

Let us now consider the part of the action that depends on $\phi_n$, Eq.~(\ref{Sn2}).  Integrating by parts and 
using the boundary conditions, we find that the contribution of the lower boundary cancels out and we obtain 
\beq
 S_n = \frac{1}{2\pi^2} \int_{{\cal B}_1} \sqrt{-g^{(3)}} \, d^3 y \lmk \frac12 \phi_n \partial_\perp \phi_n + \xi K \phi_n^2 \rmk = \frac{a_1}{2} \chi_n (\eta_1) \del_\perp \chi_n (\eta_1).
 \label{S_m}
\eeq
Then the wave function for $\chi_n$ becomes
\beq
\psi_n(\chi_{n1})\propto\exp\left(-\frac{1}{2}R_n\chi_{n1}^2\right),
\label{psinR}
\eeq
where 
\beq
R_n=-\frac{i}{N}\frac{{\dot\chi}_{n1}}{\chi_{n1}}.
\label{Rnchi}
\eeq
With $N\eta=i\tau$, this is the same as Eq.~(\ref{R_npm}) that we obtained using the WDW formalism.  Since the condition $R_n >0$ is satisfied at $\eta_1 \to-\infty$, it is guaranteed to be satisfied for all $\eta_1$.  Thus we conclude that the path integral formalism with appropriate boundary terms in the action gives the same wave function as the WDW equation with 
tunneling boundary conditions.  In both approaches the scalar field fluctuations are suppressed.

\subsection{Boundary term as the initial wave function}
\label{sec:PI2}

The new boundary term that we introduced in Eq.~(\ref{SB}) can be written as
\beq
{\tilde S}_{Bn}\equiv -\frac{1}{4\pi^2}\int_{{\cal B}_0} \sqrt{-g^{(3)}} \, d^3 y h_n\phi_n^2 =\frac{i}{2} n \chi_{n0}^2, 
\eeq
where $\chi_{n0}=\chi_n(\eta_0)$.  This term allows an interesting interpretation, which we shall now discuss.

Let us first show that the scalar field path integral in Eq.~(\ref{Psiaphi}) can be expressed as
\beq
\psi_n(\chi_{n1})\propto  \int {\cal D} \chi_n e^{i {\tilde S}_n [\chi_n]} \psi_{n0} (\chi_{n0}),
\label{intchi}
\eeq
where
\beq
 \psi_{n0} (\chi_{n0}) \equiv  e^{i{\tilde S}_{Bn}} =  e^{- n \chi_{n0}^2/2} 
 \label{initialPsi}
\eeq
and ${\tilde S}_n [\chi_n]$ is the action (\ref{Sn2}) with only $\xi K$ boundary terms included.  The integration in Eq.~(\ref{intchi}) is to be taken over paths $\chi_n(\eta)$ starting at $\chi_n(\eta_0)=\chi_{n0}$ and ending at $\chi_n(\eta_1)=\chi_{n1}$; in other words this path integral assumes Dirichlet boundary conditions.  We assume also that the functional measure includes an integral over $\chi_{n0}$.

Substituting $\phi_n=\chi_n/a$ and 
\beq
R=\frac{6}{a^2}\left(1+\frac{\ddot a}{N^2 a}\right)
\label{R}
\eeq
in the action (\ref{Sn2}) we obtain
\beq
{\tilde S}_n = \frac{1}{2} \int_{\eta_0}^{\eta_1} d\eta \left[\frac{1}{N}{\dot\chi}_n^2-N n^2\chi_n^2-N m^2 a^2 \chi_n^2 -\frac{1}{N} \frac{d}{d\eta}\left(\frac{{\dot a}}{a} \chi_n^2\right)\right] +\frac{1}{12\pi^2} \int_{\cal B} \sqrt{-g^{(3)}} \, d^3 y \frac{K}{a^2} \chi_n^2.\nn
\label{SnK}
\eeq
With $\int_{\cal B} \sqrt{-g^{(3)}} \, d^3 y = 2\pi^2 a^3$ and $K = \pm 3{\dot a}/(Na^2)$, we find that the result of integration of the total derivative in (\ref{SnK}) cancels out with the boundary term, so the result is
\beq
{\tilde S}_n = \frac{1}{2} \int_{\eta_0}^{\eta_1} d\eta \left(\frac{1}{N}{\dot\chi}_n^2-N n^2\chi_n^2-N m^2 a^2 \chi_n^2 \right).
\label{SnK2}
\eeq

The functional integral in Eq.~(\ref{intchi}) is Gaussian, so the saddle point approximation is exact.  Integrating by parts   and using the classical equation of motion for $\chi_n$, we can express the action (\ref{SnK2}) as
\beq
 {\tilde S}_n = \frac{1}{2 N} \chi_{n1} \dot{\chi}_{n1} - \frac{1}{2 N} \chi_{n0} \dot{\chi}_{n0}. 
 \label{tildeS}
\eeq
Extremizing $i {\tilde S}_n [\chi] + \ln[\psi_{n0} (\chi_{n0})]$ with respect to $\chi_{n0}$, we find 
\beq
 \dot{\chi}_{n0}= i n N \chi_{n0}, 
 \label{init1}
\eeq
which is precisely the Robin boundary condition (\ref{bcchi}).  Also, from Eqs.~(\ref{tildeS}), (\ref{initialPsi}), and (\ref{init1}), 
the amplitude (\ref{intchi}) is given by 
\beq
 \psi_n (\chi_{n1}) \propto e^{i \chi_{n1} \dot{\chi}_{n1}/ 2N}, 
\label{psichi1}
\eeq
where the second term in \eq{intchi} has cancelled out with $\Psi_0 (\chi_{n0})$. 
The combination $i {\tilde S} + \ln ( \Psi_0)$ is now finite in the limit $\eta_0 \to -\infty$, because of the cancellation. 
Eq.(\ref{psichi1}) is equivalent to Eqs.~(\ref{psinR}),~(\ref{Rnchi}) that we derived in Sec.~\ref{sec:PI2-1}.  Thus we conclude that Eq.~(\ref{intchi}) is equivalent to the path integral with Robin boundary conditions.

Now, the form of Eq.~(\ref{intchi}) is very suggestive.  We can interpret $\Psi_0 (\chi_{n0})$ as the initial wave function for the scalar field at $\eta_0 \to -\infty$.  As suggested in Ref.~\cite{Hong:2003pe}, we can think of the tunneling wave function as describing a small initial universe that tunnels to $a\approx a_*$ after reaching the bounce point at $a_0 \ll a_*$, in the limit of $a_0 \to 0$.%
\footnote{
More precisely, the background cosmology assumed in Ref.~\cite{Hong:2003pe} included a small amount of radiation with density $\rho_r = \epsilon_r / a^4$. The bounce point $a_0$ then depends on $\epsilon_r$, and the limit $a_0 \to 0$ is obtained at $\epsilon_r \to 0$. 
}
The wave function (\ref{initialPsi}) is that for a massless scalar field in the state of 
Euclidean vacuum, which is defined by requiring that the mode functions are regular at $\tau \to \infty$. 
It was shown in Ref.~\cite{Hong:2003pe} that the same quantum state is obtained if one considers a small initial universe that tunnels through a barrier in the limit when the size of the initial universe goes to zero. 
In this limit the mass of the field $\chi$ can be neglected in the wave function (\ref{psichi1}). 

\section{Conclusions}

We discussed three different approaches to defining the tunneling wave function of the universe $\Psi_T$.   The first approach is to impose the outgoing wave and regularity conditions in superspace.  This has been previously studied in Refs.~\cite{Vilenkin:1987kf,Vachaspati:1988as,Hong:2003pe}, with the conclusion that the resulting wave function is uniquely defined and describes a universe nucleating with the scalar field in a de Sitter invariant Bunch-Davies state.  The regularity condition, requiring that the absolute value of the wave function decreases with growing amplitude of scalar field fluctuations, is a non-local condition on $\Psi_T$.  Here we showed that it is equivalent to the requirement that the scalar field modes $\phi_n$ satisfy a (local) Robin boundary condition at $a\to 0$.

Our main focus in this paper was to explore the conjecture made in Refs.~\cite{Vilenkin:1984wp} that $\Psi_T$ can also be expressed as a Lorentzian path integral taken over histories interpolating between a vanishing 3-geometry $(a=0)$ and a given configuration $\{a,\phi_n\}$.  We showed that the Robin boundary conditions for $\phi_n$ require an addition of a new boundary term to the scalar field action and that the path integral is then identical to the wave function specified by the tunneling boundary conditions.  

We showed also that the path integral with the new boundary term can be expressed as a transition amplitude from a universe of vanishing size with a scalar field in the state of Euclidean vacuum.  All three approaches give identical wave functions with well behaved scalar field fluctuations, contrary to earlier claims in the literature.

Our discussion in this paper was limited to a de Sitter minisuperspace model with a scalar field included as a perturbation. 
A natural extension of this model would be to consider non-perturbative minisuperspaces, including a few degrees of freedom, but allowing large variations of the scalar field and large deviations from de Sitter geometry.  Such models with a homogeneous scalar field \cite{Vilenkin:1987kf} and with a Bianchi-IX metric \cite{delCampo:1989hy} have been studied in the framework of boundary conditions in superspace, with the conclusion that the tunneling and regularity conditions determine a unique wave function with well-behaved fluctuations.  Extension of the path integral approach to non-perturbative models remains an open problem for future research.

\begin{acknowledgments}
We are grateful to Juan Diaz Dorronsoro, Jonathan J. Halliwell, James B. Hartle, Thomas Hertog, Oliver Janssen, Jean-Luc Lehners, and Yannick Vreys for their useful and stimulating comments on the manuscript.  This work was supported in part by the National Science Foundation.
\end{acknowledgments}

\appendix

\section{Minimally coupled scalar field}

The tunneling wave function in a de Sitter minisuperspace with a minimally coupled massless scalar field was discussed in Refs.~\cite{Vilenkin:1987kf,Vachaspati:1988as}.  In this case, 
Eqs.~(\ref{H_n}) and (\ref{omega_n}) are replaced by
\beq
 {\cal H}_n = \frac{\hbar^2}{2a^2} \frac{\del^2}{\del \phi_n^2} 
 - \frac{1}{2a^2} \omega_n^2 (a) \phi_n^2 
 \label{H_n2}
 \\
 \omega_n^2(a) = (n^2-1) a^4 + m^2 a^6. 
 \label{omega_n2}
\eeq
In the wavefunction (\ref{psi}), we replace $R_n^\pm \chi_n^2$ by $R_n^\pm \phi_n^2$. 

As before, in the under-barrier range $a<a_*$ we introduce the Euclidean time $\tau$ by Eq.~(\ref{dadeta}); then the functions $R_n^\pm(a)$ can be represented by a single function $R_n(\tau)$ satisfying  
\beq
 a^2 \frac{dR_n}{d\tau}  - R_n^2 + \omega_n^2 (\tau)  = 0, 
\eeq
This can be reduced to a linear equation by the substitution 
\beq
 R_n(\tau) = - \frac{a^2}{\varphi_n} \frac{d \varphi_n}{d \tau},
\label{R_npm2}
\eeq
where the mode functions $\varphi_n(\tau)$ satisfy 
\beq
 \frac{d^2 \varphi_n}{d \tau^2} 
 + \frac{2}{a} \frac{d a}{d \tau} \frac{d \varphi_n}{d \tau} 
 - \frac{\omega_n^2}{a^4} \varphi_n = 0. 
\label{varphieq}
\eeq 
Changing the variable as $\varphi_n = \nu_n / a$, 
we rewrite the equation as 
\beq
 \frac{d^2 \nu_n}{d \tau^2} - \lkk n^2 + \lmk m^2 - 2 H^2 \rmk a^2 \rkk \nu_n = 0,
 \label{nueq}
\eeq
where we used \eq{R} and $R = 12H^2$. 

Since this equation is symmetric with respect to the replacement of $\tau \to -\tau$, 
the mode function can be written as a superposition of a symmetric function $g_{sn}(\tau)$ and an anti-symmetric function $g_{an}(\tau)$, 
\beq
 \nu_n(\tau) = A_n [g_{sn} (\tau) - g_{an} (\tau)] + B_n [ g_{sn} (\tau) + g_{an} (\tau) ], 
 \label{nu_n sym2}
\eeq
where $g_{sn} (\tau) = g_{sn}(-\tau)$ and $g_{an} (\tau) = - g_{an} (-\tau)$. 

In the limit of $\tau \to \pm \infty$, $a(\tau)\propto e^{\mp\tau}$, the solution of (\ref{nueq}) is given by 
\beq
 \nu_n(\tau) \approx  A_n e^{- n \tau} + B_n e^{n \tau}, 
\eeq
and
\beq
 R_n(\tau) \approx 
 a^2 \lmk n \frac{A_n-B_n e^{ 2 n \tau}}{A_n+B_n e^{ 2 n \tau}} \mp 1 \rmk. 
\eeq
This can be positive or zero at $\tau \to \infty$ only if $B_n = 0$. This corresponds to the boundary condition 
\beq
 \frac{d\nu_n}{d \tau} = - n \nu_n ~~~ (\tau \to \infty).
\label{bc-2}
\eeq 
It then follows from (\ref{nu_n sym2}) that 
\beq
 \frac{d\nu_n}{d \tau} = - n \nu_n ~~~ (\tau \to - \infty).
\label{bc+2}
\eeq 
These have the same form as Eqs.~(\ref{bc-}) and (\ref{bc+}).

One can easily generalize the discussion in Ref.~\cite{Hong:2003pe} and show that 
the regularity condition for $R_n^-$ ($R_n^+$) is satisfied everywhere under the barrier if it is satisfied at 
$a\to 0$ $(a=a_*)$ 
and if $\omega_n^2$ is positive everywhere under the barrier.%
\footnote{
The condition $\omega_n^2 > 0$ may not be satisfied if the field has a tachyonic mass ($m^2 <0$). 
In this case, we may need a special treatment for the homogeneous mode $(n=1)$; see Ref.~\cite{Vilenkin:1987kf}. 
}
As a result, what we need to impose is the regularity condition for $R_n^-$ at $a\to 0$,
which can be realized by either boundary condition, (\ref{bc+2}) or (\ref{bc-2}).

Turning now to the path integral formalism, most of the analysis in Sec.~\ref{sec:PI2-1} still applies.
Using Eq.~(\ref{dperp}) and $i\tau=N\eta$, we can express the boundary conditions (\ref{bc+2}) as
\beq
\partial_\perp \varphi_n = - i (n+1) a^{-1} \varphi_n  ~~~ (\tau \to - \infty) .  
\label{BC-minimal}
\eeq
A comparison with Eq.~(\ref{bc}) then shows that we need to add to the action a boundary term of the form (\ref{SB}) with $\xi=0$ and 
\beq
h_n = -i (n+1) a^{-1}. 
\eeq
As before, the lower boundary contribution to the scalar field action cancels out and Eq.~(\ref{S_m}) gives
\beq
 S_n = \frac{1}{4\pi^2} \int_{{\cal B}_1} \sqrt{-g^{(3)}} \, d^3 y \varphi_n \partial_\perp \varphi_n . 
\eeq
Then the wave function for $\varphi_n$ becomes
\beq
 \psi_n (\varphi_{n1}) \propto e^{iS_n} = \exp\left(\frac{i a_1^2}{2N} \varphi_{n1} \dot{\varphi}_{n1}\right), 
\label{psiphi2}
\eeq
which is the same as
\beq
\psi_n(\varphi_{n1})\propto\exp\left(-\frac{1}{2}R_n\varphi_{n1}^2\right)
\label{psinR2}
\eeq
obtained using the WDW formalism.

As in Sec.~\ref{sec:PI2}, the path integral over $\varphi_n(\tau)$ can be expressed as
\beq
\psi_n(\varphi_{n1})\propto  \int {\cal D} \varphi_n e^{i {\tilde S} [\varphi_n]} \psi_{n0} (\varphi_{n0}),
\label{intphi2}
\eeq
where now the action ${\tilde S}_n$ does not include any boundary terms,
\beq
 \psi_{n0} (\varphi_{n0}) = e^{-(n+1) a_0^2 \varphi_{n0}^2/2}. 
\eeq
and the integration is over histories $\varphi_n(\tau)$ satisfying Dirichlet boundary conditions at $\eta_0\to -\infty$ and $\eta_1$.  Following the same steps as in Sec.~\ref{sec:PI2}, one can show that the result is the same as in Eq.~(\ref{psiphi2}).

\bibliography{reference}

\end{document}